\documentclass[preprint,amssymb, numerical, aip, pof]{revtex4-1}
\usepackage{graphicx}
\usepackage{amsmath}
\usepackage{color}
\usepackage{graphics}
\usepackage[USenglish]{babel}
\usepackage{varioref}
\usepackage[T1]{fontenc}
\usepackage{ae}
\usepackage{fancyhdr}
\usepackage[latin1]{inputenc}
\usepackage{subfigure}
\usepackage{appendix}
\usepackage{psfrag}

\usepackage{xcolor}
\usepackage{amssymb}
\usepackage{newtxtext}
\usepackage{newtxmath}
\usepackage{hyperref}

\usepackage{geometry}
\begin{document}

\title{Superhydrophobic surfaces to reduce form drag in turbulent separated flows}
\author{J.-P. Mollicone}
\affiliation{Department of Mechanical Engineering, Faculty of Engineering, University of Malta, MSD 2080, Msida, Malta}
\email[]{jean-paul.mollicone@um.edu.mt}
\author{F. Battista}
\affiliation{Department of Mechanical and Aerospace Engineering, Sapienza University of Rome, via Eudossiana 18, 00184, Rome, Italy}
\author{P. Gualtieri}
\affiliation{Department of Mechanical and Aerospace Engineering, Sapienza University of Rome, via Eudossiana 18, 00184, Rome, Italy}
\author{C.M. Casciola}
\affiliation{Department of Mechanical and Aerospace Engineering, Sapienza University of Rome, via Eudossiana 18, 00184, Rome, Italy}
\date{\today}%

\begin{abstract}
The drag force acting on a body moving in a fluid has two components, friction drag due to fluid viscosity and form drag due to flow separation behind the body. When present, form drag is usually the most significant between the two and in many applications, streamlining efficiently reduces or prevents flow separation. As studied here, when the operating fluid is water, a promising technique for form drag reduction is to modify the walls of the body with superhydrophobic surfaces. These surfaces entrap gas bubbles in their asperities, avoiding the direct contact of the liquid with the wall. Superhydrophobic surfaces have been vastly studied for reducing friction drag. We show they are also effective in reducing flow separation in turbulent flow and therefore in reducing the form drag. Their conceptual effectiveness is demonstrated by studying numerical simulations of turbulent flow over a bluff body, represented by a bump inside a channel, which is modified with different superhydrophobic surfaces. The approach shown here contributes to new and powerful techniques for drag reduction on bluff bodies.
\end{abstract}


\maketitle

\section{Introduction}
\label{sec:intro}

One of the current global challenges is to reduce the amount of energy used for human 
activities in order to preserve natural resources and safeguard the environment. 
A considerable amount of energy is used to overcome drag forces and much effort 
is dedicated to efficient control and reduction drag.
Two drag sources exist in a turbulent flow over a bluff, non-lifting body moving in an incompressible 
fluid (note: for lifting bodies the induced drag needs to be taken into account,
while wave drag concurs in compressible and in free-surface flows in presence of gravity). 
The two sources of drag are: friction at the wall associated with fluid viscosity 
(friction drag) and lack of pressure recovery behind the body associated with flow separation
(form drag) \cite{landau1987fluid}.

When present, flow separation prevails, since friction drag is subdominant at large speed. 
In such conditions, the traditional approach to drag control still mostly relies on body 
streamlining, i.e. body shape optimisation aimed at preventing/limiting flow separation. 
The drag force $D$ is expressed by the drag coefficient, $C_D = D/(1/2 \rho U_\infty^2 L^2)$, 
in terms of fluid density $\rho$, free stream velocity $U_\infty$ and body frontal section area 
$A \sim L^2$, $L$ being the body characteristic length scale. While the friction drag 
coefficient decreases with the Reynolds number, $\text{Re} = \rho U_\infty L/\mu$ 
($\mu$ is the dynamic fluid viscosity), scaling as a power law with negative exponent 
($\alpha = -1/2$ for laminar cases and $\alpha \approx -1/5$ for turbulent cases), the form drag 
coefficient is Reynolds independent at high enough Reynolds numbers \cite{batchelor2000introduction}, 
being connected to the phenomenon of ``dissipative anomaly'' in turbulence \cite{frisch1995turbulence}.
It is therefore wise to target the form drag in all cases where flow separation cannot be 
avoided by simple streamlining.

Reducing form drag is a challenging problem in both fundamental and applied fluid dynamics. 
Several solutions have been proposed, with both active devices (e.g. blow/suction  
\cite{kametani2015effect,beck2018drag}, gas injection \cite{elbing2013scaling}) and passive 
devices (e.g. Large-Eddy Breakup Devices \cite{spalart2006direct} and riblets 
\cite{mele2020drag,li2015turbulent}). The drawback of active techniques is the energy necessary 
to actuate the control mechanisms and the complexity of the actuator system. On the other hand, 
passive devices lose their effectiveness in off-design conditions.

A richer set of possibilities are available for friction drag control. Among these, relevant 
for our present purposes, are passive techniques based on superhydrophobic surfaces (SHSs)  
\cite{rothstein2010slip,schellenberger2016water} and on liquid impregnated 
surfaces \cite{fu2017liquid}. Their application, though limited to applications in water, 
shows encouraging results in 
friction drag reduction 
\cite{rastegari2019drag,park2013numerical,park2014superhydrophobic,min2004effects,rastegari2015mechanism} 
in both laminar \cite{lee2016superhydrophobic,gruncell2013simulations} 
and turbulent regimes \cite{daniello,aljallis2013experimental} 
and have been studied in a variety of domains such as, for example, 
channels \cite{jelly2014turbulence}, Taylor-Couette flow \cite{srinivasan2015sustainable},  
pipe flow \cite{costantini2018drag} and around rotors \cite{choi2019wake}.
Inspired by nature \citep{wang2015bioinspired}, SHSs trap gas bubbles in their asperities, 
allowing liquid moving onto the gaseous phase to slip, unlike at solid walls where the 
liquid adheres \cite{turk2014turbulent,hu2017significant,rajappan2019influence,im2017comparison,lee2018effects,xiong2017influence}.

The simulations discussed in the present paper suggest that SHSs can also be effective in 
reducing form drag of bluff bodies in turbulent flow. 
Turbulence models cannot reliably predict the flow behaviour outside the 
range of conditions for which the model has been developed and tested. Since this is the case 
of SHSs, the full Navier-Stokes equations are solved on a supercomputer by the
Direct Numerical Simulation (DNS) approach \cite{moin1997tackling}, which resolves 
all the relevant physical scales of the flow (no turbulence models).

The configuration addressed here follows from a compromise between fidelity to the actual 
conditions and numerical efficiency. The geometry consists of a bump (or bulge, 
representing the bluff body) mounted on one of the two otherwise planar and parallel surfaces 
bounding a channel of height $2h$. The bump, 
of characteristic length $2h$, 
features a system of SHSs with the purpose to investigate their effect on the flow separation 
occurring behind the obstacle and the related form drag. In absence of SHSs, the bump is known 
to make the flow separate, creating a strong shear layer emanating from the separation point 
and producing a recirculating region \cite{stella2017scaling,krankdirect,schiavo2017turbulent}. 
The configuration, with no SHSs has been extensively investigated using classical statistical 
tools of turbulence theory \cite{mollicone2018turbulence,mollicone2017effect,passaggia2018optimal,
Kahler_2016,matai2019large,fadla2019investigation}.
\begin{figure}
\centering
\includegraphics[height=0.25\textwidth]{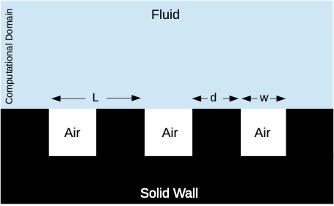}
\includegraphics[height=0.25\textwidth]{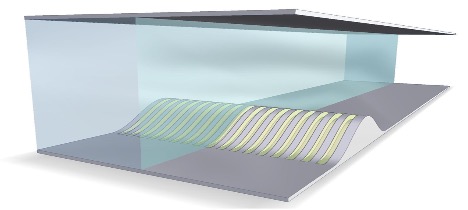}
\caption{\label{fig:sketch} The superhydrophobic surface made by alternating stripes 
and grooves (left panel), is reproduced in numerical simulations by no-slip (gray) and 
free-shear (yellow) patterns (right panel).}
\end{figure}
\section{Case Setup}

Direct Numerical Simulations (DNSs) of turbulent channel flow containing a bump at one 
of the walls are performed by solving the incompressible Navier-Stokes equations, 
\begin{align}
\nabla\cdot \vec{u} &= 0\\
\frac{\partial \vec{u}}{\partial t} + \nabla \cdot \vec{u} \otimes \vec{u} &= - \nabla p +\nu \nabla^2 \vec{u} + \vec{f}
\label{eq:NS}
\end{align}
where $t$ is the time, $\vec{u}$ is the fluid velocity vector, $p$ is the hydrodynamic pressure, 
$\nu$ is the kinematic viscosity and $\vec{f} = \Delta p/L_x \hat{e}_x$ is the constant pressure 
gradient that sustains the flow rate inside the channel, ($\hat{e}_x$ is the unit vector in the 
streamwise direction). The solver used is Nek5000~\citep{nek5000}, an open-source code based 
on the spectral element method (SEM)~\citep{patera1984spectral}. The SEM combines high accuracy 
of spectral methods and the flexibility, in terms of geometrical configuration, of finite 
element approaches. The computational domain has dimensions 
$[L_x\times L_y \times L_z] = [20\times 2 \times 2\pi] \times h$, where $x$, $y$, and $z$ are 
the streamwise, wall-normal and span-wise coordinates respectively and $h$ is the nominal half 
channel height. All quantities are made dimensionless with respect to the bulk velocity 
$U_b=Q/(2h)$, where $Q$ is the flow rate per unit width. Flow is in the $x$ direction with 
periodic boundary conditions in both $x$ and $z$ directions. The periodicity in the 
streamwise direction replicates a periodic array of bumps, similar to the experimental 
configuration found in \cite{Kahler_2016}. The periodic configuration is instrumental in 
avoiding spurious effects that artificial inflow/outflow boundary conditions could induce and 
allows the analysis of an almost isolated bump, with definite flow reattachment and negligible 
streamwise correlation. The simulations are carried out at bulk Reynolds numbers 
$\text{Re}=5000$. The maximum friction Reynolds number, achieved close to the bump tip for the 
reference simulation is $\text{Re}_\tau=u_\tau h/\nu=540$, where $u_\tau=\sqrt{\tau_w/\rho}$ is 
the local friction velocity and $\tau_w$ is the shear stress at the wall. The friction velocity 
and the viscous length, $y_\tau=\nu/u_\tau$, which are the characteristic parameters in 
turbulent wall-bounded flows, are used as reference quantities to normalise the velocity and 
length scales, respectively, and are indicated with the $^+$ superscript. To the best of 
our knowledge, the friction Reynolds number is the highest reached in the literature concerning 
DNS of similar geometries.

No-slip boundary conditions are enforced at the top and bottom walls except on the bump, where 
the superhydrophobic surface is modelled by alternating no-slip/shear-free boundary conditions,
see figure~\ref{fig:sketch}. These boundary conditions mimic the presence of ridges of width 
$d$ (solid wall) alternated with grooves of width $w$. Physically, gas is assumed to be stably 
trapped in the grooves, hence allowing the slippage of the liquid on top, corresponding to 
vanishing shear force, \cite{liravi2020comprehensive}. The pattern is aligned in the streamwise 
direction, see right panel of figure~\ref{fig:sketch}. A liquid-gas interface is pinned at the 
edge of the grooves, resembling a stable Cassie-Baxter state, as shown in the left panel of 
figure~\ref{fig:sketch}. The no-slip and the no-penetration boundary conditions are enforced on 
the ridge, being a solid wall, and the shear-free and the no-penetration boundary conditions are 
enforced on the liquid-gas interface. The interface is then a fixed boundary on which a 
perfect slip condition is applied.  The groove width, $w$, is changed whilst the solid fraction, 
i.e. the solid surface to the overall surface ratio, is kept constant at $\phi_S=0.5$, 
implying that $d=w$. The periodicity $L$ therefore changes, since $L=d+w$.

\begin{table}
\caption{ Simulation details: simulation name (first column), periodicity length of the 
no-slip/free-shear stripes normalised with half channel height (second column), periodicity 
length of the stripes normalised with the viscous length (third column), grid spacing 
normalised with the viscous length in streamwise (forth column), wall normal (fifth column) 
and spanwise (sixth column) directions, number of computational points in millions 
(seventh column). The two values in the fifth column refer to the grid spacing at the 
channel center (max) and at the wall (min). }
{\begin{tabular}{lcccccc}
      Simulation  & $L$   &   $L^+$ & $\Delta x^+$ &$\Delta y^+_{max/min}$ & $\Delta z^+$ & Grid points\\[3pt]\hline
       REF   & --  & -- & 8.0 & 4/0.5  & 9.0 & 101 $\cdot$ 10$^6$\\
       L20   & 0.071 & 20 & 7.5 & 4/0.5 & 1.8 & 544 $\cdot$ 10$^6$\\
       L40  & 0.142 & 40 &  7.5 & 4/0.5 & 3.5 & 274 $\cdot$ 10$^6$\\
       L80   & 0.284 & 80 & 7.5 & 4/0.5 & 7.0 & 138$\cdot$ 10$^6$\\\hline
  \end{tabular}}
  \label{tab:1}
\end{table}

Table~\ref{tab:1} shows a summary of the simulation parameters. The first column lists the 
names of the simulations, where REF is the reference simulation with a fully no-slip bump wall 
condition and the superhydrophobic cases are labelled depending on the dimension of the grooves. 
The second column shows the periodicity length normalised with the half channel height whilst 
the third column lists the periodicity length normalised with the viscous length. 
The fourth, fifth and sixth columns provide the computational grid spacing normalised with the 
viscous length in streamwise, wall-normal and spanwise directions, respectively. The grid 
is uniform in the streamwise and spanwise directions and stretched along the wall normal 
direction to cluster grid nodes toward the walls. The number of nodes in the spanwise direction 
are increased with decreasing periodicity to obtain adequate grid resolution over the 
no-slip/shear-free boundary conditions on the bump. The seventh column lists the number of 
collocation points. The simulations have been carried out on supercomputing facilities. 
The high computational costs are due to the fine computational grids 
required to resolve the turbulent scales and the SHS pattern especially when the periodicity 
length is small. The statistics discussed are based on 300 independent samples of the flow 
field collected at instants separated in time by more than the flow turnover time which 
largely exceeds the maximum correlation time of velocity and pressure fluctuations. 
Convergence is enhanced by exploiting the statistical spanwise homogeneity of the flow. 
The channel and bump dimensions and the Reynolds number are kept fixed and therefore the 
periodicity length is the control parameter of the system. 
Separation and reattachment points are 
evaluated by finding the streamwise position where the wall-normal derivative of the mean 
velocity parallel to the wall is zero. To evaluate the reattachment point, the velocity 
component parallel to the wall coincides with the streamwise velocity, whilst for the 
separation point, the mean velocity tangential to the bump is considered.

\begin{figure}
\centering
\includegraphics[height=0.9\textwidth]{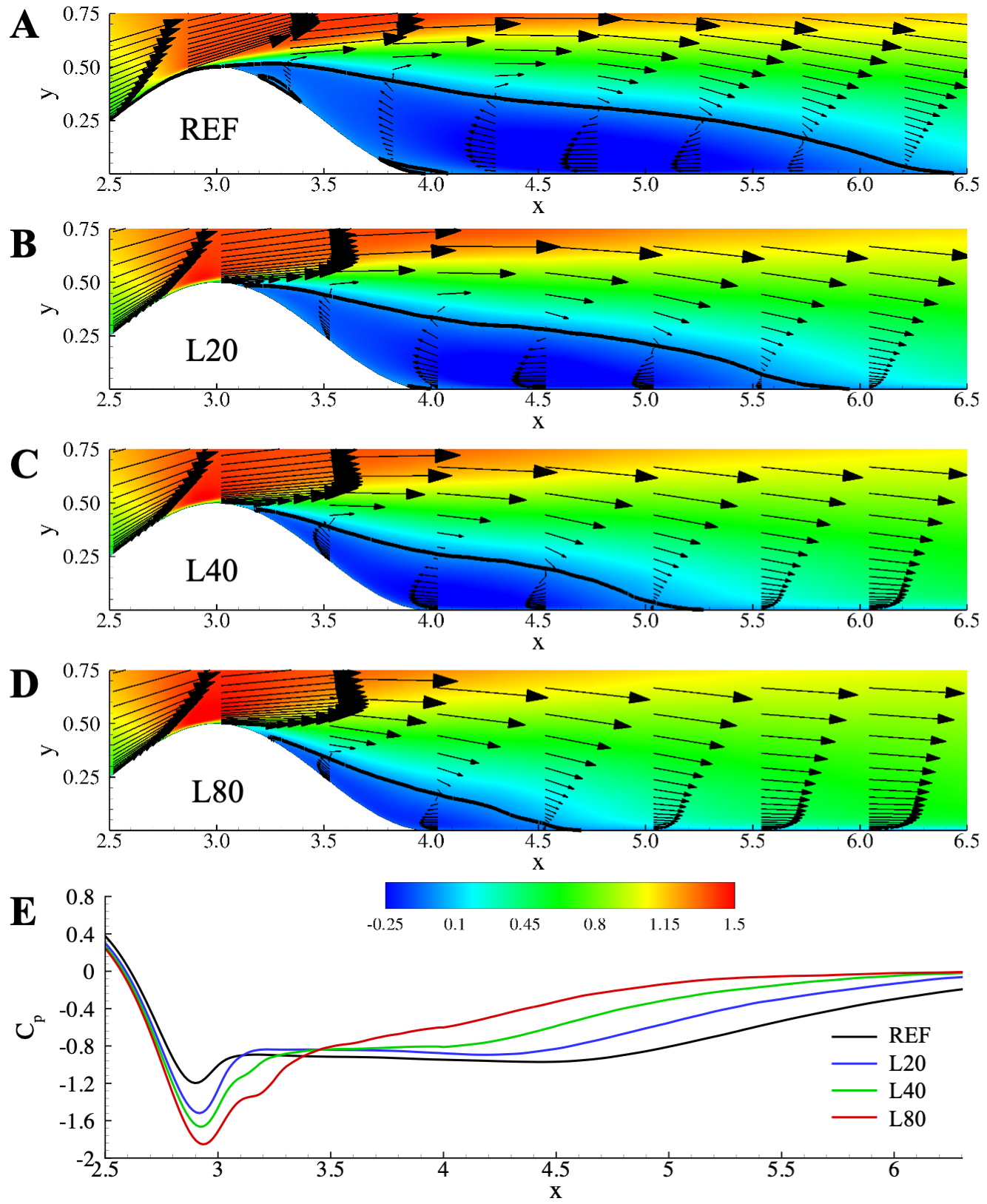}
\caption{\label{fig:ux_med} Coloured contour of mean streamwise velocity normalised with the 
bulk velocity and corresponding zero isoline (black solid line), in panels from A to D for 
simulations REF, L20, L40, and L80, respectively. Pressure coefficient along the bottom wall 
for all the simulations in panel E.}
\end{figure}

\section{Results}

\begin{figure}
\centering
\includegraphics[width=0.49\textwidth]{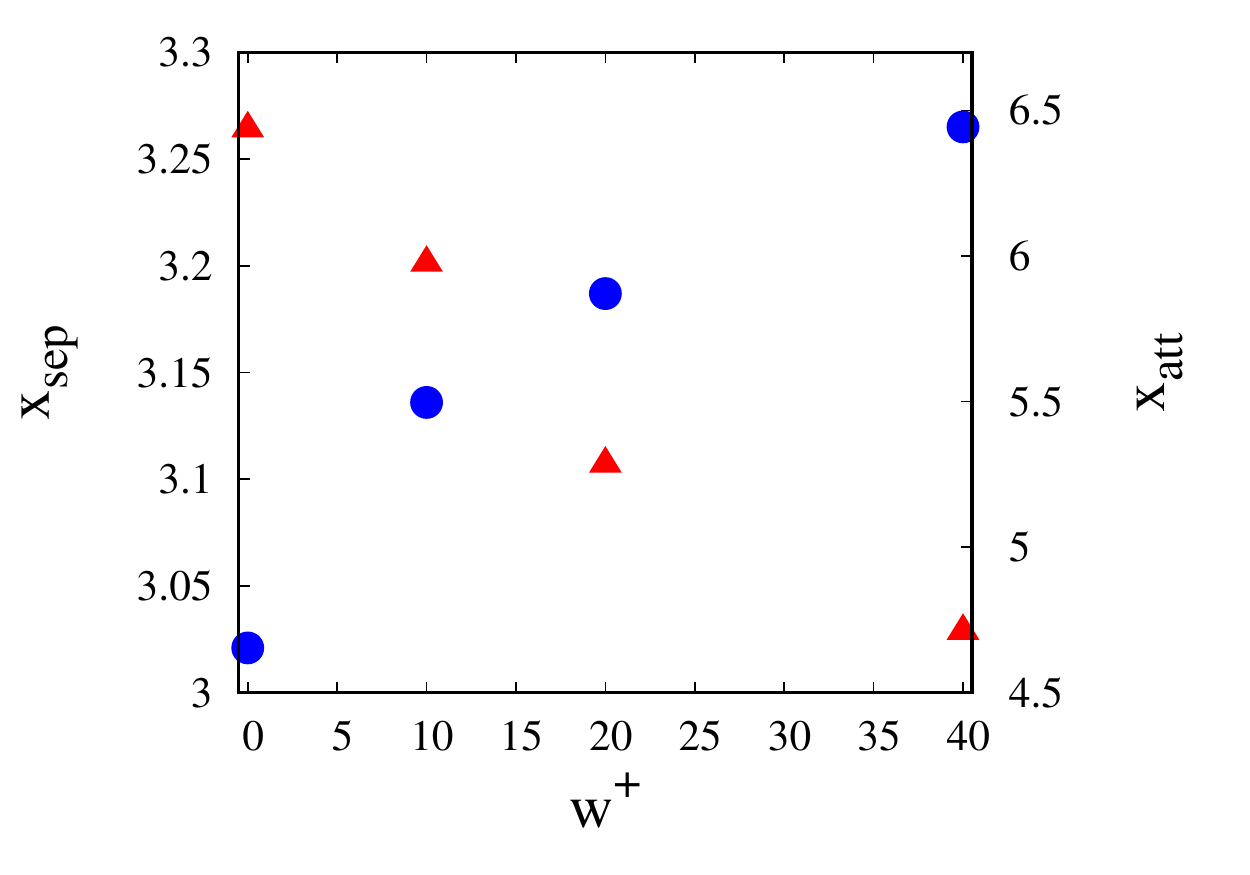}
\includegraphics[width=0.49\textwidth]{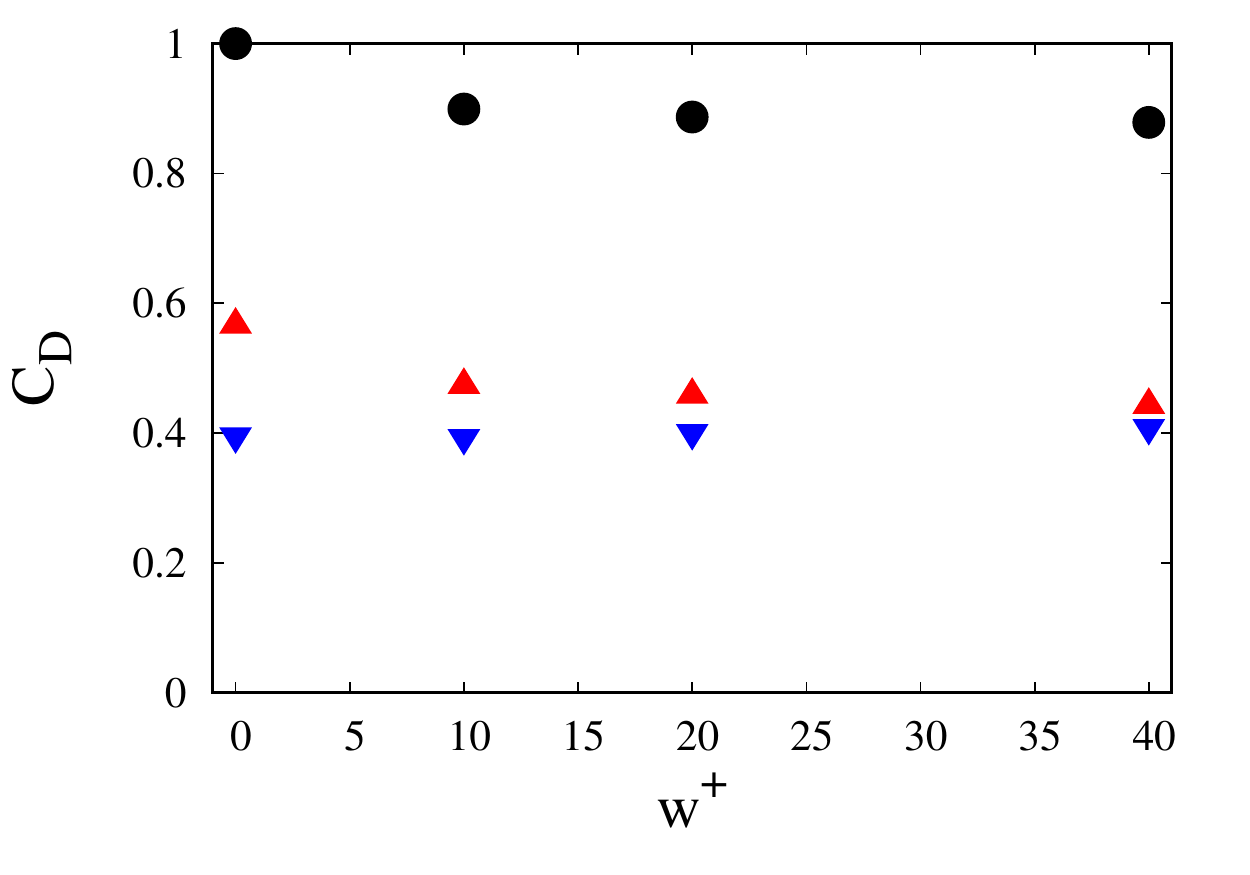}
\caption{\label{fig:cd_sep} Left panel: plot of separation (blue circle
\textcolor{blue}{$\medbullet$}) and reattachment points (red triangle 
\textcolor{red}{$\blacktriangle$}) against the groove width normalised with the viscous length.
Right panel: plot of drag coefficient components; total 
(black circle $\medbullet$), form (red triangle \textcolor{red}{$\blacktriangle$}) and 
friction (blue gradient \textcolor{blue}{$\blacktriangledown$}),  against the groove 
width normalised with the viscous length. 
}
\end{figure}

One of the two central results of this work is illustrated in figure~\ref{fig:ux_med}, panels 
A to D. In the plots, the arrows reproduce the mean flow velocity averaged on 300 instantaneous 
flow fields sampled along the simulation and separated by the velocity autocorrelation time. 
The extension of the recirculation region behind the obstacle decreases with increasing the 
periodicity length $L$ of the stripes at $d/w=1$.
The left panel in figure~\ref{fig:cd_sep} shows 
that when the groove dimension increases, the separation point $x_{sep}$ moves downstream 
and the reattachment point $x_{att}$ upstream, resulting in a smaller separation bubble.
The change in the mean velocity in figure~\ref{fig:ux_med} is accompanied by a change 
in the pressure distribution along the bump, which is reproduced in panel E.
As expected, when reducing the size of the recirculating region, 
the pressure behind the obstacle increases, therefore the form drag decreases, 
see eq.~\eqref{eq:cdfd}. 

The decrease in form drag is the second central result of the work and is shown in the 
right panel of figure~\ref{fig:cd_sep}, which concerns the drag exerted on the obstacle. 
The drag coefficient is defined as 
\begin{align}
C_d = -\frac{2}{L_x} \int_{walls} \langle\vec{t}\rangle \cdot \hat{e}_x dl \, ,
\label{eq:cd}
\end{align}
while the friction and form drag coefficients are defined as 
\begin{align}
C_d^{form} &= -\frac{2}{L_x} \hat{e}_x \cdot \int_{walls} P \vec{n} \, dl \\
C_d^{friction} &= -\frac{2}{L_x} \hat{e}_x \cdot \int_{walls} \mu \frac{\partial \langle 
\vec{u}_{||}\rangle} {\partial \vec{n}} \, dl \, ,
\label{eq:cdfd}
\end{align}
respectively. $\vec{t}$ is the (dimensionless) traction at the wall and $\hat{e}_x$ 
is the unit vector in the streamwise direction, the angular brackets define the ensemble 
average in time and spanwise direction, $P$ is the average pressure, $\vec{n}$ is the unit 
normal exiting the fluid domain, and the dot represents the scalar product.
The total drag is reduced by up to 12\% with respect to the reference case with no SHS, 
since the form drag is reduced by up to 22\%. 

The decrease in recirculation bubble dimension is due to the 
increase in turbulent velocity fluctuations induced by the wall pattern.
The wall pattern induces vortical structures that are anchored to the no-slip/shear-free interfaces 
and that transfer fluctuations on to the no-slip surface and away from the 
wall~\citep{costantini2018drag,jelly2014turbulence,turk2014turbulent}.
The increase in fluctuations promotes the delay of the 
separation point and consequently an earlier reattachment of the flow, resulting in a sensible 
decrease of the form drag.

\begin{figure}
\centering
\includegraphics[width=1.0\textwidth]{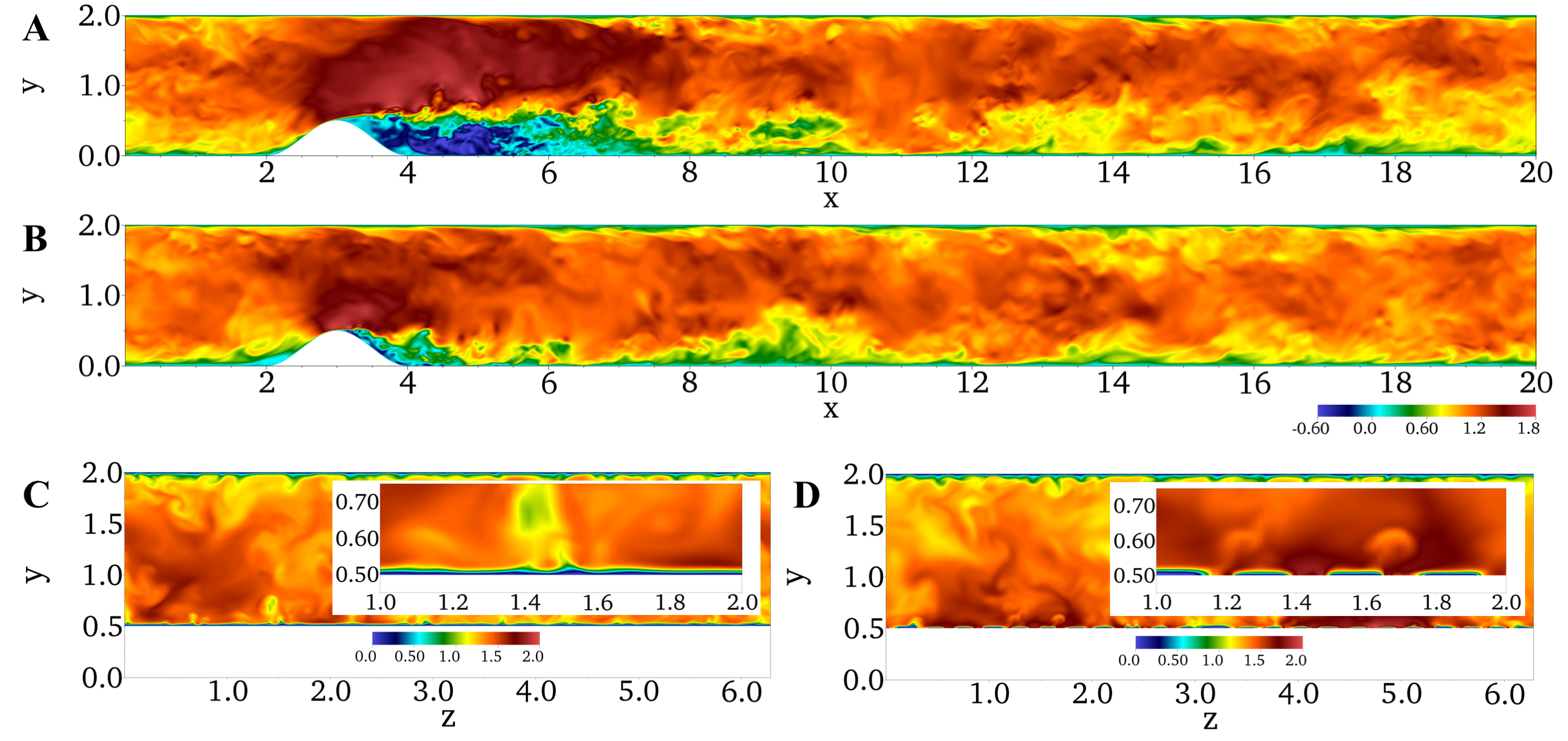}
\caption{\label{fig:inst} Instantaneous streamwise velocity, normalised with the bulk velocity, 
in $x-y$ plane (panels A and B) and $y-z$ plane at the top of the bump (panels C and D) 
concerning reference simulation (A and C) and simulation L80 (B and D); the insets in panel C 
and D report a detail of the wall, with the latter showing the superhydrophobic surface.}
\end{figure}

To convey an idea on the turbulence fluctuations, figure~\ref{fig:inst} shows a configuration 
of the instantaneous streamwise velocity in the reference REF case and case L80. Panels A and B 
show a longitudinal $x-y$ section whilst panels C and D show the transverse $y-z$ plane at the 
tip of the bump with a magnification near the bump wall in the inset. The instantaneous 
fields in the longitudinal section confirm that the separation region seen in the reference 
case, panel A, almost disappears in the superhydrophobic case, panel B. In the transverse 
$y-z$ plane, the effect of the SHS, in an alternating zero/finite velocity at the bottom wall, 
results in modified turbulence structure at the wall, panel D.

As anticipated, the decrease of the recirculation bubble dimension is induced by the increase 
of the turbulent velocity fluctuations. The analysis can be made quantitative by considering 
the turbulent kinetic energy (TKE) equation,
\begin{align}
\frac{\partial \Phi_j}{\partial x_j} = -\varepsilon + \Pi + \langle \frac{\Delta p'}{L_x} u_x'\rangle
\label{eq:tke}
\end{align}
where $\varepsilon = \langle 1/{\text{Re}}  (\partial u'_i/\partial x_j) (\partial u'_i/\partial x_j) \rangle $ is 
the TKE pseudo-dissipation rate, 
$\Pi = \langle u'_i u'_j\rangle {\partial U_i }/{ \partial x_j}$ is the production of TKE, 
$\langle {\Delta p' u_x'}/{L_x} \rangle$ is the external source of energy due to the 
fluctuating part of the pressure gradient-velocity correlation, and $\Phi_j$ is the spatial 
flux contributing zero net power when integrated over the whole domain given by
\begin{align}
\Phi_j = U_j k + \frac{1}{2} \langle u_i' u_i' u_j'\rangle +\langle p' u_j'\rangle 
-\frac{1}{\text{Re}} \frac{\partial k}{\partial x_j}
\label{eq:tke_flux}
\end{align}
where $k=1/2 \langle u'_i u'_i \rangle$ is the turbulent kinetic energy. The energy provided by the fluctuations of 
inlet/outlet pressure drop is negligible with respect to the other source term, 
$\Pi$, i.e. $\langle {\Delta p' u_x'}/{L_x} \rangle \simeq 10^{-5} \Pi$.

\begin{figure}
\centering
\includegraphics[width=0.8\textwidth]{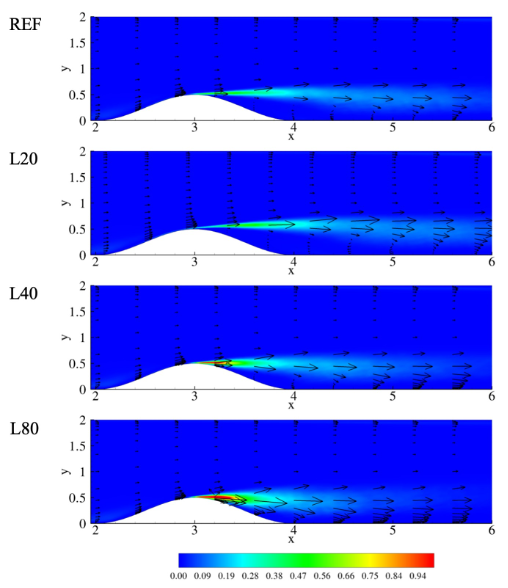}
\caption{\label{fig:tke_budget} Turbulent kinetic energy production $\Pi$ as background colour 
and spatial energy flux $\Phi_j$ as vectors.
}
\end{figure}

Figure~\ref{fig:tke_budget} reports the production 
term $\Pi$ (background colour) and the energy spatial fluxes $\vec{\Phi}$ (vectors) for 
the different cases considered in the manuscript. 
The production term increases when the periodicity length increases since the superhydrophobic 
surfaces produce higher velocity fluctuations with respect to the classical no-slip 
wall, as previously discussed. Whilst this increase of velocity fluctuations would 
produce an increase of the friction drag, overwhelmed by the fluid slip at the wall, it 
produces a sensible decrease of the form drag by delaying the separation. The 
production term feeds the flow close to the bump 
wall~\cite{mollicone2017effect,mollicone2018turbulence} with additional energy 
with respect to the reference case resulting in a delayed separation point and 
a smaller recirculation bubble.

\section{Conclusions}

Superhydrophobic surfaces (SHS) reduce the form drag by reducing the 
separation bubble behind a bluff body in turbulent flow. The separation point is delayed whilst the 
reattachment point is anticipated, corresponding to a decrease of the separation bubble 
dimensions of up to 35\%. The culprit of this phenomenon is the substantial modification
of the turbulent kinetic energy production mechanisms induced by the superhydrophobic surface.
The turbulent kinetic energy production increases in the shear layer, and spatial fluxes are able to 
sustain an higher level of turbulent fluctuations near the wall. It follows that the resulting 
flow can withstand more effectively the adverse pressure gradient delaying the 
separation point, reducing the dimensions of the separation bubble and facilitating pressure 
recovery in the reattachment region. In the present analysis, the main assumption is that 
the gas entrapped in the SHSs asperities is stable and no liquid infusion in these asperities 
occurs. We also assume a non-deformable interface, which might be sometimes different from reality.
Dedicated experimental studies are needed to address the conditions at which the 
gaseous bubbles stably remain in the asperities and the interface can be assumed stable or not. 

\section*{Author's Contribution}
All authors contributed equally to this work.

\section*{Acknowledgements}
We acknowledge the CINECA award under the ISCRA initiative, for the availability of high 
performance computing resources. 

\section*{Data Availability}
The data that support the findings of this study are available from the corresponding author
upon reasonable request.

\bibliography{main}

\end{document}